\documentclass[preprint,prl,amsmath,amssymb]{revtex4}

\usepackage{graphicx}
\usepackage{dcolumn}
\usepackage{bm}

\def\jon{j_{\rm on}}
\def\joff{j_{\rm off}}
\def\tauoff{\tau_{\rm off}}
\def\kT{k_{{}_{\rm B}}T}
\def\G{\Omega}
\def\tauD{\tau_{{}_{\rm D}}}

\begin{document}

\title{Non-equilibrium raft-like membrane domains under continuous recycling}
\author{Matthew S. Turner${}^{1}$, Pierre Sens${}^{2}$ and Nicholas D. Socci${}^3$}
\affiliation{${}^1$Department of Physics, University of Warwick, Coventry CV4 7AL, UK}
\affiliation{${}^2$ESPCI, 10 Rue Vauquelin, 75231 Paris Cedex 05, France}
\affiliation{${}^3$Computational Biology Center, Memorial Sloan-Kettering Cancer Center, 1275 York Avenue, NY, NY 10021.}
\date{\today}
\pacs{
82.39.-k 
87.15.Kg 
87.15.Rn 
87.16.Dg 
}
\keywords{recycling, membrane, coarsening, raft, domain}

\begin{abstract}
We present a model for the kinetics of spontaneous membrane domain (raft) assembly that includes the effect of membrane recycling ubiquitous in living cells. We show that the domains have a broad power-law distribution with an average radius that scales with the 1/4 power of the domain lifetime when the line tension at the domain edges is large. For biologically reasonable recycling and diffusion rates the average domain radius is in the tens of nm range, consistent with observations. This represents one possible link between signaling (involving rafts) and traffic (recycling) in cells. Finally, we present evidence that suggests that the average raft size may be the same for all {\it scale-free} recycling schemes.
\end{abstract}
\maketitle
Liquid-liquid phase separation in mixed membranes is now a well known phenomenon. Separated domains in vesicles have been observed by fluorescence labelling and the size of circular sphingomyelin (SM)-enriched domains can reach almost half that of a 20$\mu$m-sized vesicle \cite{baumgart,kahya,kellerprl}. These domains quickly reassumed their circular shape if deformed, and merged with each other to create circular domains, all phenomena consistent with the existence of strongly immiscible liquid domains. There has been much recent biological interest in ``lipid rafts'' which are believed to be liquid ordered membrane microdomains containing certain proteins and enriched in glycosphingolipids and cholesterol \cite{raft1,raft2,edidinreview}. These are thought to have many important functions, including in signal transduction and in the sorting of proteins. Experimental estimates of the size of lipid rafts {\it in vivo} are in the few tens to 100 nm range \cite{pralle,dietrich,simonsreview,rao,keller} or even smaller \cite{fret}. We refer to these as ``intermediate'' sized domains in what follows to distinguish them from the micron scale (or larger) domains observed in reconstituted membranes that are close to equilibrium.
%

We first review why intermediate sized domains are not expected to form on two component membranes at equilibrium in the absence of long range interactions \cite{seul} or recycling \cite{edidin}. We assume that the surface coverage of domains remains small and write a simple Flory-Huggins model for the grand potential energy per area as a function of the distribution of the sizes of domains, defined by the dimensionless concentration $c_{n}$ of domains containing $n$ monomers. All concentrations are per monomeric area $s=\pi(b/2)^{2}$ with the monomer diameter perhaps $b\sim5$nm if we identify the effective monomeric unit {\it in vivo} as a typical raft-resident proteins together with its associated lipid ``skirt'' \cite{vereb}. 
\begin{equation}
\G=\sum_{n=1}^{\infty} c_{n}\left(\log\frac{c_{n}}e -\mu n +\gamma\sqrt{n}\right)
\label{FH}
\end{equation}
Here and below all energies are measured in units of $\kT$. A line tension $\gamma/(2\sqrt{\pi s})$ acts at the perimeters of all the domains. The chemical potential $\mu$ is conjugate to the total area fraction of domains $\phi=\sum_{n=1}^{\infty} n\>c_{n}$. Thermodynamic equilibrium then corresponds to
\begin{equation}
c_{n}=e^{-\gamma\sqrt{n}+\mu n}
\label{ceq}
\end{equation}
The average domain size is $\bar n=\phi/N$ with $N=\sum_{n=1}^{\infty} c_{n}$ the total density of domains of all sizes. When $\phi$ is very small $\mu(\phi)\sim\log\phi$ is large and negative and almost all domains are very small (monomeric) in size. For $\mu=0$ there exists a {\it critical} total area fraction $\phi_{c}=\sum_{n=1}^{\infty} ne^{-\gamma\sqrt{n}}$, again made up of small domains provided $\gamma\gg1$. There is a phase transition for $\phi>\phi_{c}$ beyond which the area fraction of finite-sized, essentially monomeric, domains $\phi_{c}$ remains constant but the distribution becomes highly bimodal with a few very large domains accounting for the remainder $\phi-\phi_{c}$ \footnote{If instead $\gamma\lesssim 1$ the behaviour is physically rather uninteresting, approaching a gas of non-interacting, primarily monomeric, domains for all $\phi\ll 1$.}.  This is at odds with many experimental observations and hence motivates a non-equilibrium model, involving recycling, which will occupy the remainder of this paper.
\section{Non-equilibrium: recycling}\label{theory}
The evolution of the domain size distribution on an infinite membrane may be written via the following master equation
\begin{eqnarray}
{\frac{dc_{n}}{dt}}&=\sigma(n)+\sum_{m=1}^{\infty} k_{n,m} \>c_{n+m}-k'_{n,m}\>c_{n}c_{m}\hskip1cm{} \nonumber\\ &+\frac12\sum_{m=1}^{n-1}k'_{m,n-m}\>c_{n-m}c_{m}-k_{m,n-m}\>c_{n}
\label{master}
\end{eqnarray}
Here $\sigma(n)$ controls the lipid recycling, as will be discussed in further detail below; $\sigma=0$ when there is no recycling. The kernals $k_{n,m}$ and $k'_{n,m}$ control, respectively, the rates of domain scission in which one domain of $n+m$ monomers breaks into two, of size $n$ and $m$, and domain fusion, in which two domains containing $n$ and $m$ monomers fuse to form a single domain of size $n+m$, see Fig~1a. A similar approach has been rather successful in describing the kinetics of wormlike micelles \cite{Tur90}. 

We assume that two domains fuse whenever they diffuse into contact. Thus $k'=D/s$ with $D$ a characteristic diffusion constant for the domains and $s$ the area of the smallest (monomeric) domains.  The characteristic, microscopic ``diffusion'' time scale $\tauD=1/k'=s/D\approx 10^{-5}$s for biological membranes. We propose to neglect any size-dependence of the diffusion constant of the domains, which is a fair approximation \cite{saffman}. This simplifies the analytic analysis and should only weakly affect our results, which rely primarily on the fact that the recycling rate is much slower than $1/\tauD$. 

The scission kernal $k_{n,m}$ is now determined by the principle of detailed balance, an approach that is appropriate provided the longest intra-domain relaxation time following a fusion or scission event is shorter than the domain collision time. This should be satisfied on average for fluid domains \cite{DLOvsDLD} where the rate of subsequent inter-domain events should then converge to that found at equilibrium for each (pair of) domain(s), in spite of the fact that the size distribution may be far from equilibrium. By inspection of Eq~(\ref{master}) together with Eq~(\ref{ceq})
\begin{equation}
k_{n,m}=e^{-\gamma(\sqrt{n}+\sqrt{m}-\sqrt{n+m})}k'_{n,m}
\label{k}
\end{equation}
where $k'_{n,m}=k'=1/\tauD$ is treated as a constant in what follows. This kinetic scheme, without recycling, will yield asymptotic domain growth reminiscent of the coarsening of crystal domains \cite{lifshitzslyozov}.

In order to analyse the effect of lipid recycling we first consider the ``monomer deposition / raft removal''  (MDRR) recycling scheme in which raft lipids and proteins are brought to the membrane at random as single `monomer' sized units with a rate $\jon$ and entire rafts are lost from the membrane at random with a rate $\joff$, irrespective of their size, see Fig~1b. Thus $\sigma(n)$ appearing in Eq~(\ref{master}) takes the form
\begin{equation}
\sigma(n)=\jon\delta_{n,1}-\joff c_{n}
\label{sigmamonomerraft}
\end{equation}
with $\delta_{i,j}$ the usual Kronecker delta. It is easily shown that $\phi=\jon/\joff$ at steady state. In general Eq~(\ref{master}) requires numerical solution, see Fig~2 and \footnote{See EPAPS Document No. [...] for movies showing the evolution of the size distribution.}, although an asymptotic solution is possible in the most interesting regime $\gamma\gg 1$, as discussed below.

It can be seen from Fig~2 that the domain size distribution is broad, indeed there is significant contribution to the total area fraction from domains with $n\lesssim\bar n^{2}$. The distribution depends only weakly on $\gamma$ when $\gamma$ is itself large because all domain scission events are then rare.

To investigate this `scissionless' large $\gamma$ regime further we proceed by neglecting all scission terms in Eq~(\ref{master}) and obtain
\begin{equation}
{\frac{dc_{n}}{dt}}=\sigma(n)-k'N\>c_{n} 
+k'\frac12\sum_{m=1}^{n-1} c_{n-m}c_{m}\label{asymptoticmaster}
\end{equation}
where we will later have to check our solutions for self-consistency, which will translate to establishing a lower bound for $\gamma$.  

The equations $\dot c_{n}=0$ for $n=1,2\dots$ can then be used to build up $c_{n}$ recursively. For MDRR recycling given by Eq~(\ref{sigmamonomerraft}) this yields
\begin{equation}
c_{n}=A_{n}(\jon\tauD)^{n}/(\joff\tauD+N)^{2n-1}\label{termn}
\end{equation}
with $A_{n}=(2n-2)!/(2^{n-1}n!(n-1)!)$. Eq~(\ref{termn}) still involves $N$ and so the size distribution is only explicitly determined after solving for $N=\sum c_n$.  The resulting $c_n$ appears as the solid curve on Fig~2. In the most biologically relevant regime \footnote{$\phi\approx0.1$, $\tauD\approx 10^{-5}$s and $\joff\gtrsim100$s \cite{vereb} might be typical.} $\phi/(\joff\tauD)\gg 1$ and $N=\sqrt{2\phi\joff\tauD}$. Eq~\ref{termn} then approaches
\begin{equation}
c_{n}\approx \sqrt{\phi\joff\tauD/(2\pi)}\>n^{-3/2}\quad{\rm for}\quad1\ll n \ll {\bar n}^{2}
\label{capproxasymp}
\end{equation}
which appears as the dashed curve on Fig~2. The average domain area in this regime is
\begin{equation}
\bar n=\sqrt{\phi/(2\joff\tauD)}
\label{nbarasymp}
\end{equation}
as shown in Fig~3 (dashed line).

When scission is rare the average domain radius is found to be $R=10$--$70$nm, for $b=1$--5nm respectively for recycling with rate $\joff=10^{-2}{\rm s}^{-1}$, see Fig~3. As can be seen from the raft area Eq~(\ref{nbarasymp}) the raft radius scales only with the 1/4 power of the recycling rate or diffusion time and hence is in the tens to $\sim 100$nm range for all reasonable physiological values. Restoring dimensionality the threshold line tension $\gamma^{\star}\approx10$ is about $0.6\kT$/nm when $b=5$nm. The line tension between different membrane phases may have several origins. Hydrophobic thickness mismatch \footnote{AFM measurements on artificial bilayers report \cite{simonsreview,riniaAFM} report mismatches of order $10\%$ between liquid ordered and disordered domains.} gives rise to line tensions that depend upon the elastic properties of the membrane and are in the $\kT$/nm range. Other, more directly chemical or ordering-related, incompatibilities may also contribute. Thus the asymptotic results of this section, in which we assumed that the domains rarely break, may have wide applicability and full numerical solution of the master equation may not always be necessary.

The scission rate Eq~(\ref{k}) has a maximum at $m=1$ corresponding to shedding single monomers. For domains with $n\gg 1$, such as those of the average size, the maximal rate is $k_{1,n-1}\approx e^{-\gamma}k'$ which can be used to establish a self-consistency condition: Only if the number of monomeric scission events in the typical residence time of a domain is much less than $\bar n$ will the scission rate be negligible. The product of this rate and the lifetime $\joff^{-1}$  is much less than $\bar n$ whenever $\gamma\gg\gamma^\star$ with $\gamma^{\star}=\frac12\log{\frac2{\phi\joff\tauD}}$.

Another intuitive `scale free' recycling scheme is the ``monomer deposition / monomer removal''  (MDMR) scheme in which raft lipids and proteins are again assumed to be brought to the membrane at random as small `monomer' sized units with a rate $\jon$ but domain material (raft lipids and proteins) are removed from the membrane in monomeric units, at random, with a rate $\joff$, irrespective of the size of the raft on which they reside. While perhaps of less biological relevance this scheme, together with MDRR, form the most extreme examples of an entire class of possible recycling schemes: they correspond to the removal of monomers, through to whole rafts, respectively and involve no characteristic size scales for the recycling. As such it would be surprising if any intermediate `partial raft'  recycling scheme produced behavior that was not bounded qualitatively by the two extremes that we consider. Within this scheme 
\begin{equation}
\sigma(n)=\jon\delta_{n,1}-\joff(n c_{n}-(n+1)c_{n+1})
\label{sigmamonomermonomer}
\end{equation}
with $\phi=\jon/\joff$ at steady state, as before. The general solution of Eq~(\ref{master}) is again obtained numerically, see Fig~4. 

Fig~5 shows that the mean domain size is the same for both of these recycling schemes, suggesting that this is true for {\it all} scale-free recycling schemes. To understand this note that the lifetime of every domain monomer is the same under both schemes. It is possible to propose other schemes that are not scale-free but rather have intrinsic size scales and it is far from clear that such schemes may not be biologically relevant. Such scales might correspond to a characteristic raft size at which endocytosis occurs. While it is straightforward to treat such schemes by an appropriate choice of $\sigma(n)$ the mechanisms by which lipids are recycled in the cell are still rather poorly understood \cite{recycling}  and a thorough investigation of the numerous possible schemes lies outside the scope of the present work.

\section{Discussion}
We have shown how the non-equilibrium nature of membranes can lead to steady-state domain sizes that are {\it intermediate} in size, typically in the tens to 100nm range for all biologically reasonable recycling rates and membrane diffusion constants. This result seems to agree well with experimental observations of lipid rafts on the plasma membranes of different cells and is marked contrast to the large domains observed on artificial membranes as they approach equilibrium. We believe that our results will be of interest to those working to realise membrane recycling in model (bio)chemical systems as well as those seeking candidates models that may provide a better understanding of lipid rafts in living cells.

It is now being realized that signaling and traffic in cells may be closely related processes. Our model shows how the regulation of membrane traffic (recycling) might simultaneously control the raft sizes. It is also quite plausible that the size of rafts controls certain aspects of their function and this would give a direct connection between signaling and traffic.

We are currently studying the potentially important effects of finite (cell) membrane area, which may result in a failure of the mean field approach for the largest rafts. Also of interest is the response of the raft size distribution to perturbations in the recycling, such as a step change in the monomer deposition rate. This will provide a biologically accessible way of relating response time(s) to this model and hence the rate of the underlying recycling mechanism.


\newpage
FIGURE CAPTIONS\\
\\
\noindent
Fig~1. The model: (a) all domains can undergo scission and fusion (b) domains are deposited and removed from the membrane stochastically.\\
\\
Fig~2. The steady state domain size distribution for MDRR recycling with $\phi=\jon/\joff=0.1$, $\joff\tauD=10^{-3}$ and $\gamma=2$ ($\diamond$) 4 ($\triangle$) and 6 ($\square$). The mean domain contains 1.3, 2.3 and 5.3 monomers respectively, rather small because a fast recycling rate was chosen for numerical convenience. The asymptotic variation for large line tensions Eq~(\ref{termn}) is shown (solid line) as is Eq~(\ref{capproxasymp}) (dashed line) appropriate for  sizes $1<<n<<\bar n^{2}$.\\
\\
Fig~3. The variation of the mean domain size $\bar n$ and radius $R$ with $\phi/(\joff\tauD)$ (solid line) assuming that the domain scission rate is negligible and $b=5$nm. The power law variation for large, but physiologically relevant, $\phi/(\joff\tauD)$ is also shown (dashed line) as is the average residence time of the domains $\tauoff$, with $\tauD=10^{-5}$s and $\phi=0.1$.\\
\\
Fig~4. The steady state domain size distribution for the MDMR recycling scheme. The same values as Fig~2 are used except $\gamma=3$  ($\diamond$) 6 ($\triangle$) and 9 ($\square$). The mean domain contains 1.5, 5.2 and 7.0 monomers respectively, larger than in Fig~2 because of the larger line tension $\gamma$. The asymptotic behavior for the MDRR scheme at large line tensions Eq~(\ref{termn}) is shown (solid line), as is Eq~(\ref{capproxasymp} (dashed line), but it overestimates the very small number density of the largest domains present under this scheme.\\
\\
Fig~5. The steady state mean domain size $\bar n$ as a function of $\gamma$ for both MDRR (polygons) and MDMR (crosses) recycling, with $\phi=0.1$ and $\joff\tauD=10^{-1}$ ($\times$ and $\triangle$) $\joff\tauD=10^{-3}$ ($+$ and $\square$) $\joff\tauD=10^{-4}$ (crosses and pentagons). The asymptotic values of $\bar n$ for $\gamma\gg\gamma^{\star}$ from Fig~3 are shown as dashed lines at $\bar n= 1.4$ ($\joff\tauD=10^{-1}$), $\bar n= 7.6$ ($\joff\tauD=10^{-3}$) and $\bar n= 23$ ($\joff\tauD=10^{-4}$). The corresponding values of the crossover tension $\gamma^{\star}$ are 2.6, 5.0 and 6.1 respectively and are close to each inflection point.


\newpage
\pagestyle{empty}
\begin{figure}[h]\includegraphics{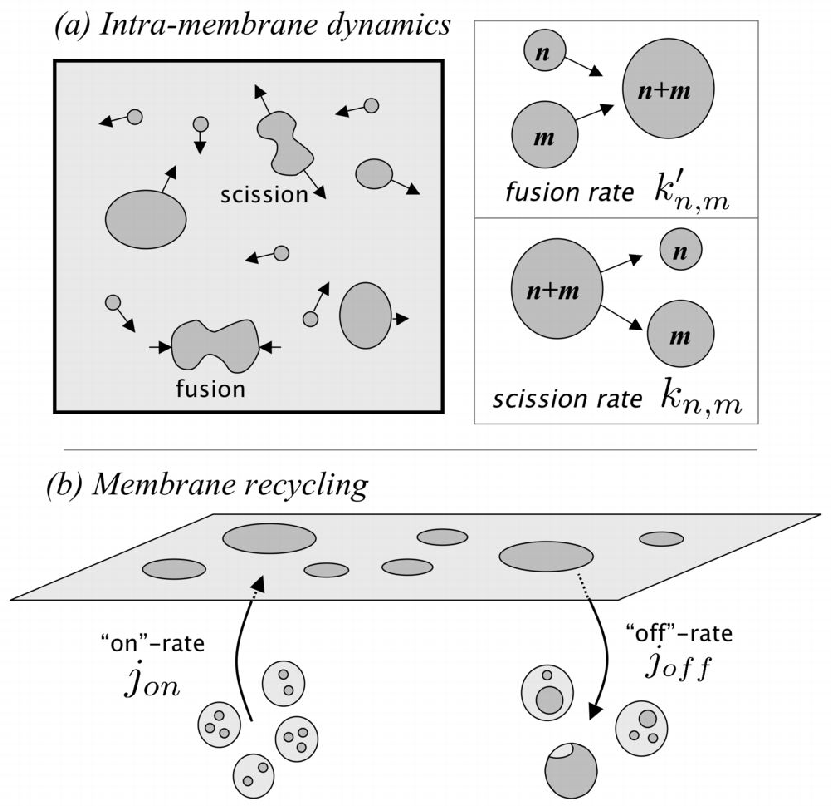}\label{kinetics}\end{figure}
\vskip3in
Fig~1

\newpage
\pagestyle{empty}
\begin{figure}[h]\includegraphics{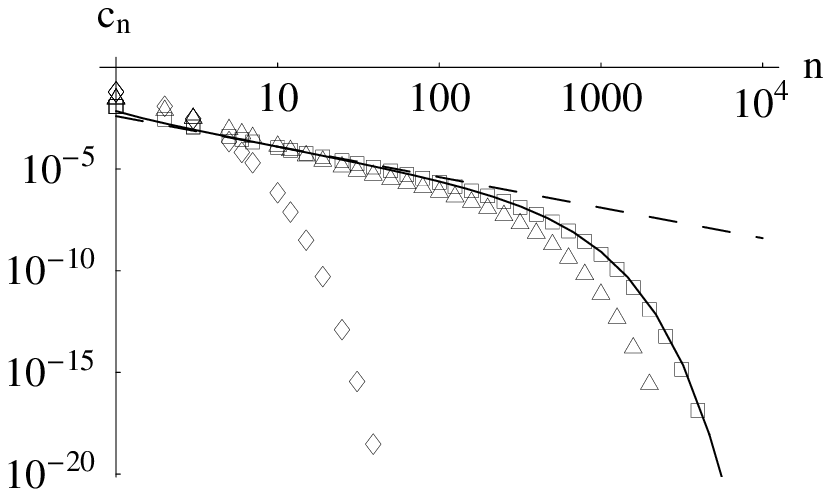}\label{numerical}\end{figure}
\vskip5in
Fig~2

\newpage
\pagestyle{empty}
\begin{figure}[h]\includegraphics{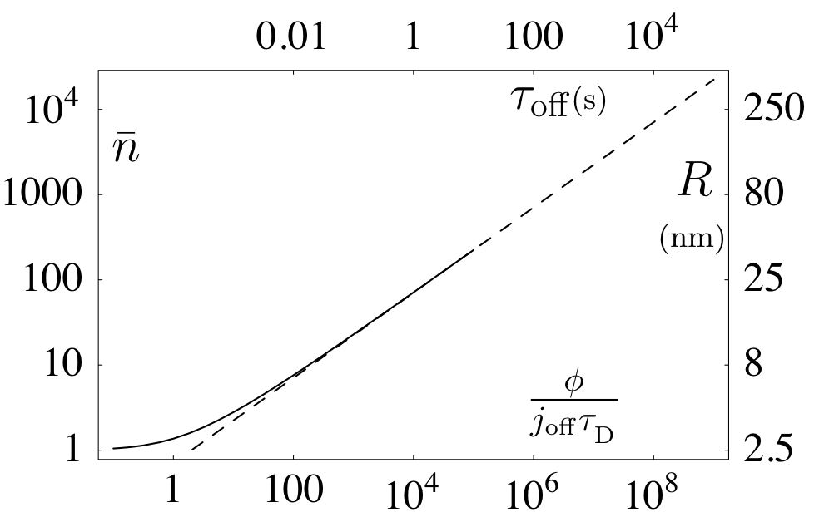}\label{lognbarofphit}\end{figure}
\vskip5in
Fig~3

\newpage
\pagestyle{empty}
\begin{figure}[h]\includegraphics{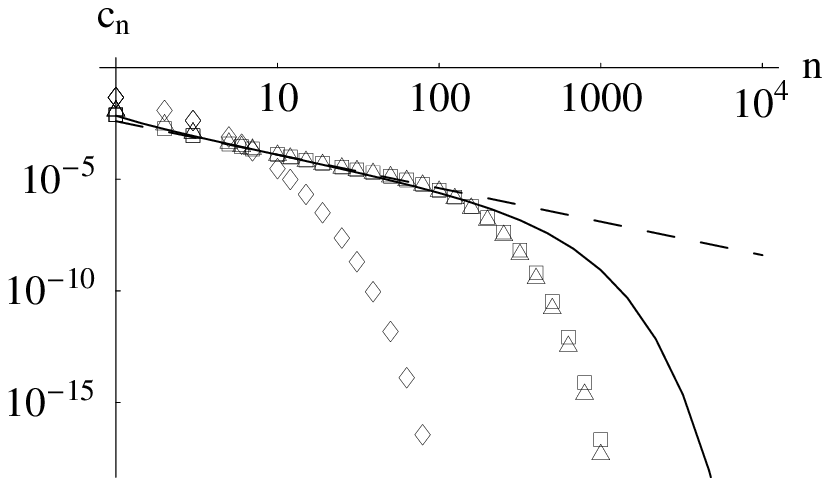}\label{numerical2}\end{figure}
\vskip5in
Fig~4

\newpage
\pagestyle{empty}
\begin{figure}[h]\includegraphics{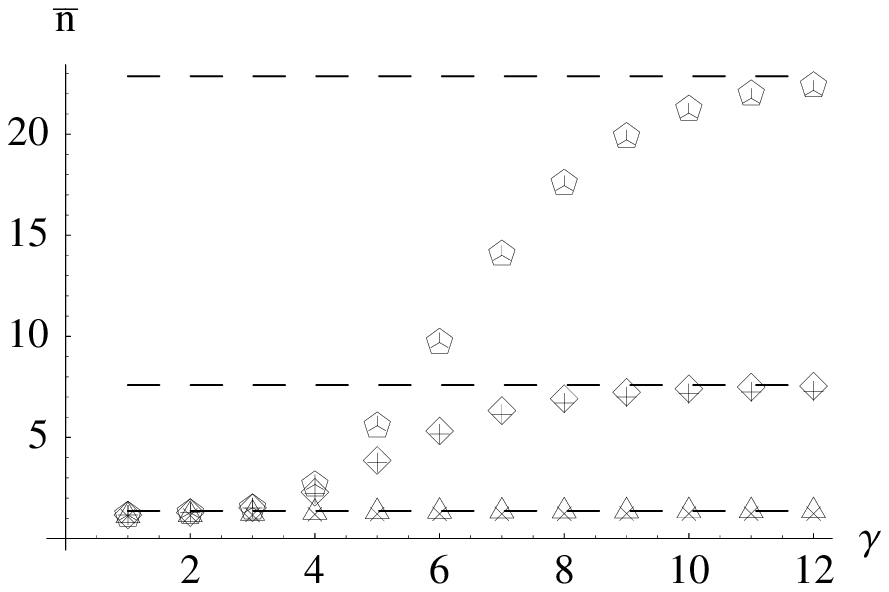}\label{nbartwoschemes}\end{figure}
\vskip3.5in
Fig~5

\end{document}